\documentclass{iopconfser}
\usepackage{graphicx} 
\usepackage{subfigure}
\usepackage{gensymb}
\usepackage{amsmath}
\usepackage{xcolor}
\usepackage{hyperref}
\usepackage{orcidlink}

\begin{document}

\title{Temperature induced optical scatter changes in titania-germania coatings}

\author{D. P. Kapasi$^{1}$\,\orcidlink{0000-0001-8189-4920}, T. Counihan$^{1}$, J.R. Smith$^{1}$\,\orcidlink{0000-0003-0638-9670}, S. Tait$^{2}$\,\orcidlink{0000-0003-0327-953X} and G. Billingsley$^{2}$\,\orcidlink{0000-0002-4141-2744}}

\affil{$^1$The Nicholas and Lee Begovich Center for Gravitational-Wave Physics and Astronomy, California State University, Fullerton, CA 92831, USA}
\affil{$^2$ LIGO, California Institute of Technology, Pasadena, CA 91125, USA}
\email{dkapasi@fullerton.edu}

\begin{abstract}
Titania doped with tantala is the high index material (high n) for the optical coatings used in LIGO and Virgo and its thermal noise limits LIGO/Virgo observations of astrophysical sources. In this paper, we study temperature induced changes to optical scatter of a multilayer highly reflective coating comprised of silica (low n) and titania doped with germania (high n) as a potential candidate to reduce coating thermal noise in ground-based observatories operating at room temperature. We observe that the scatter measured at 8$\degree$ in a small region is low, with a median starting Bidirectional Reflectance Distribution Function (BRDF) of $1.1 \times 10^{-7}\,\mathrm{str}^{-1}$ increasing to $1.2 \times 10^{-6}\,\mathrm{str}^{-1}$ through annealing. The results presented here show the potential of adopting titania doped with germania coatings for future upgrades to LIGO and Virgo and as a pathfinder coating for Cosmic Explorer, a next-generation detector.
\end{abstract}

\section{Introduction}
Direct observations of compact binary coalescences involving black holes and neutron stars, made possible by the LIGO-Virgo-KAGRA network of gravitational-wave (GW) observatories \cite{aLIGO, aVirgo, KAGRA, gw150914}, have enabled population studies of black holes \cite{gwtc3}, provided insights into the equation of state and mass distributions for neutron stars \cite{ns_mass, gw170817}, and uncovered potential new formation channels for intermediate mass black holes \cite{gw231123}. To detect GWs, the distance between four test masses (mirrors), made of fused silica with highly reflecting (HR) coatings, is accurately measured in a modified Michelson interferometer. 

The HR coatings of the LIGO and Virgo mirrors are comprised of alternating layers of silica ($\mathrm{SiO_{2}}$) and titania doped tantala ($\mathrm{TiO_{2}:Ta_{2}O_{5}}$) forming the low and high index materials, respectively \cite{granata2020}. Currently, LIGO's astrophysical reach is limited by coating thermal noise (CTN) of the test mass mirrors. In particular, $\mathrm{Ta_{2}O_{5}}$, an ingredient of the high index coating material, has high mechanical loss making it difficult to surpass the CTN limits seen in these instruments \cite{harry}. To mitigate this, investigations into other materials to replace the high index layer is ongoing \cite{mariana}. Some potential candidates for the amorphous coatings include titania doped with germania ($\mathrm{TiO_{2}:GeO_{2}}$) and silicon nitride ($\mathrm{SiN_{x}}$) \cite{vajente2021, granata2022}. 

This paper focuses on the observation of optical scatter during annealing (heat treatment) of a full HR stack (52 layers) comprised of $\mathrm{SiO_{2}}$ and $\mathrm{TiO_{2}:GeO_{2}}$ with a custom layer thickness design (LIGO sample PL2604). The coating was designed at CSU Fort Collins and deposited by Laboratoire des Matériaux Avancés (LMA) in March 2025 using ion beam sputtering  with a target refractive index of 1.897 at 1064\,nm for $\mathrm{TiO_{2}:GeO_{2}}$. The deposition process included titania grids, which are under active research as a means to reduce optical absorption. The temperature induced optical scatter was measured using the air annealing scatterometer (section \ref{section:setup}) and the results from this study are presented in section \ref{section:results}.

\section{Experimental setup}
\label{section:setup}
\begin{figure}[t!]
\centering
\includegraphics[width=1\linewidth,height=5cm,keepaspectratio]{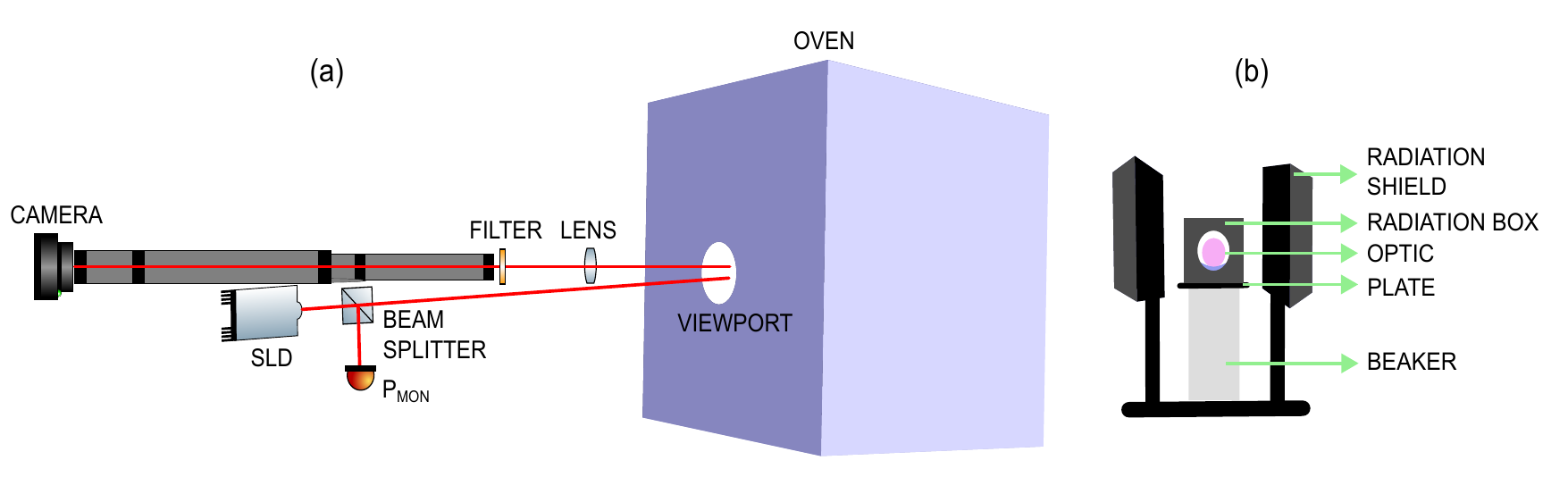}
\caption{Overview of the optical layout for the air annealing scatterometer. (a) shows the experimental schematic where a superluminiscent diode (1041\,nm) is used to probe the optic installed inside the oven. The reflected beam from the optic makes a shallow 8$\degree$ angle with respect to the incident beam, and the scatter from the optic is continuously monitored over the course of the annealing run. (b) shows the design of the optic holder inside the oven. }
\label{fig:setup}
\end{figure}

An overview of the setup is shown in Figure \ref{fig:setup}. Briefly, it consists of a superluminescent diode (SLD) at 1041\,nm which is used to probe the optical coatings \cite{elenna2021}. 10$\%$ of the power is picked-off for power monitoring ($\mathrm{P_{mon}}$). The reflected beam has a shallow 8$\degree$ angle with respect to the incident beam. This reflected light is continuously recorded throughout all stages of heat treatment using a high resolution Apogee camera ($4096 \times 4096$ pixels). In front of the camera, a narrow-band filter is installed to pass the SLD light and block thermal radiation from the oven. Inside the oven, the sample is mounted on a fused silica beaker and is surrounded with a radiation box and shield to minimize thermal radiation in the field of view of the optic. This is illustrated in Figure \ref{fig:setup}(b). Prior to installation, the optic is visually inspected for damage and  cleaned with First Contact, a polymer based solution to remove any residual particulates. The air annealing scatterometer has been discussed in more detail elsewhere \cite{rezac2023}. 

During operation, LabView software is used to control the experiment and for data acquisition. Each data cycle consists of the following steps - (a) Turn on the laser. (b) Record the process variables (temperature, time, power monitors, heater power, temperature setpoint). (c) Take a bright image (eg: 5\,s exposure time). (d) Transfer the image to storage folder. (e) Turn off the laser. (f) Record dark image for same exposure time. (g) Transfer dark image. (h) Repeat cycle. A custom Matlab script is used for analysis and details of the analysis is discussed in the following section \cite{rezac2023}. 

\section{Results}
\label{section:results}
\begin{figure}[t!]
\centering
\includegraphics[width=1\linewidth,height=10cm,keepaspectratio]{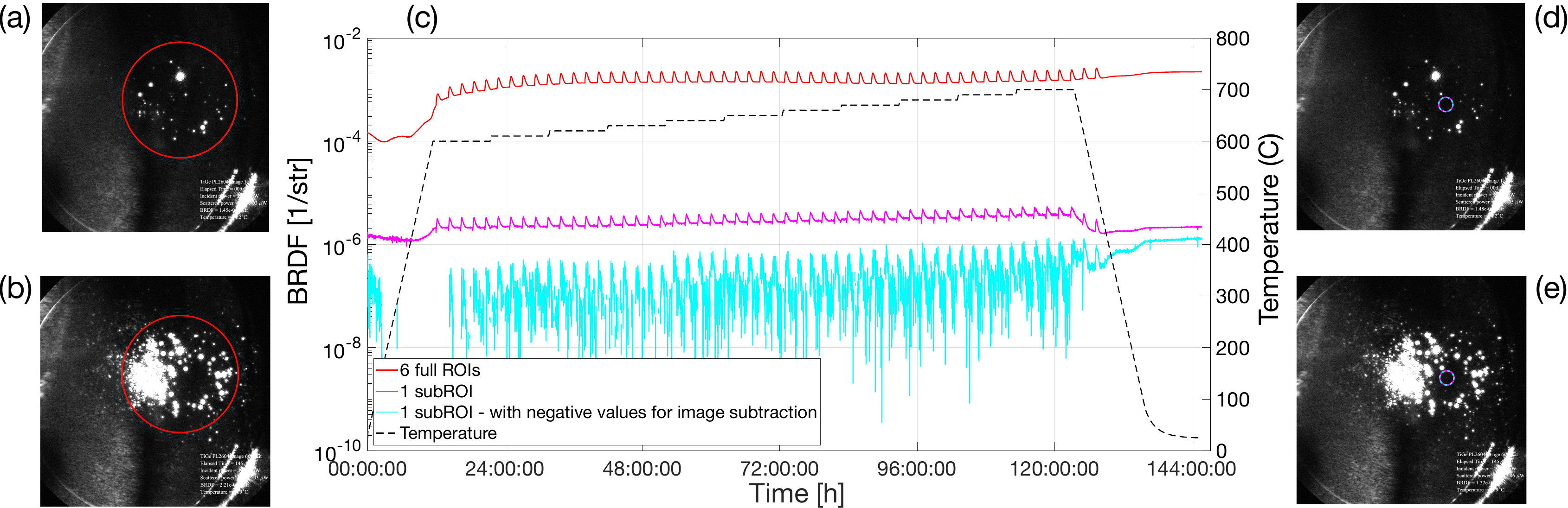}
\caption{Optical scatter from annealing of the sample. (a) and (b) show the bright images of the sample at $t=0$ and $t=end$ respectively with the same exposure time. (c) shows the temperature induced scatter from the sample. The red curve shows BRDF from the large ROI (explained in text, illustrated in (a) and (b)) and the pink curve shows BRDF from a single subROI analysis where the ROI is highlighted in (d) and (e) corresponding to the $t=0$ and $t=end$ respectively. Both the red and pink curves bias the negative values from the image subtraction to zero. The blue curve in the above graph shows the BRDF from one subROI allowing for negative values from the image subtraction process.}
\label{fig:results}
\end{figure}

The Bidirectional Reflectance Distribution Function (BRDF) is used to quantify scatter by providing information on the angular distribution of light when it reflects from any surface. Mathematically, it is written as
\begin{equation}
    \begin{centering}
        \text{BRDF} = \frac{\mathrm{scattered\,power\,per\,unit\,solid\,angle}}{\mathrm{incident\,power\,per\,unit\,projected\,area}} \approx \frac{P_s / \Omega_s}{P_i \cos \theta_s}
    \end{centering}
\end{equation}
where $P_{s}$ is the total scattered power over solid angle $\Omega_{s}$, $P_{i}$ is the total incident power and $\theta_{s}$ is the angle of scattering which in our case is $\theta_{s} = 8\degree$. The scatter results for the $\mathrm{TiO_{2}:GeO_{2}}/\mathrm{SiO_{2}}$ HR sample are shown in Figure \ref{fig:results}. Figure \ref{fig:results}(a) and Figure \ref{fig:results}(b) show images of the optic taken at $t=0$ and $t=end$ for the experimental run. To analyze the scatter, elliptical regions of interest (ROI) are defined which encompass different portions of the sample illuminated by the SLD. The innermost ROI is shown with a solid red circle in Figure \ref{fig:results}(a) and Figure \ref{fig:results}(b). The bright images are corrected for blackbody radiation from continuous heating, and later are subtracted from its respective dark image to reduce the overall noise and bias from hot pixels in the camera (if any). In the situation where the BRDF is low, i.e., the scatter within the ROI is comparable to the rest of the optic, the data from five additional ROIs (nested elliptical areas capturing data from beyond the previous ROI) is taken and analyzed to accurately determine the BRDF. The pixels within each ROI are counted and plotted against the total pixel area. The y-intercept from the linear fit gives the BRDF. This is shown as the red curve in Figure \ref{fig:results}(c). Additionally, negative values from the image subtraction are biased to be zero. The periodic spikes in the data occur every 120 minutes due to thermal glitching of the camera since two out of the four cooling fans failed during operation, thereby increasing the temperature of the CCD camera periodically. From this curve, we can see that the overall optical scatter increases by more than an order of magnitude as a function of annealing temperature. From Figure \ref{fig:results}(b), we can conclude that many bright scatterers have appeared during the experimental run. Some of the scatterers at the 9 o'clock position already start to appear during the ramp up, i.e. before the first soak at 600\,C. These scatterers may be bubbles or other damage arising in the substructure of the coating, and there is an ongoing effort within the GW community to confirm their origin.

Looking at the start (Figure \ref{fig:results}(a)) and end (Figure \ref{fig:results}(b)) images from the experimental run, we can see that there is an area within the ROI where there is not much growth in scatter. This encouraged us to quantify the scatter using a single small ROI (referred to as subROI) and the location of the subROI is shown in Figure \ref{fig:results}(d),(e). The BRDF from the subROI is shown as the pink curve in Figure \ref{fig:results}(c). We notice that the scatter in this case is better by at least 2 orders in magnitude when compared to the larger ROIs (red curve). Across all the soak temperatures, between 600\,C and 700\,C (heat treatment temperatures of interest to yield significant CTN improvement), the performance of the scatter in the subROI is $\sim$500 times lower in magnitude than the larger ROIs. This suggests that the optical coating recipe may be suitable for test masses in ground-based GW observatories operating at room temperature.

To further remove any observational bias, negative values of the image subtraction were considered while quantifying the BRDF. This is shown as the blue curve in Figure \ref{fig:results}(c). The gaps in the data indicate negative values which cannot be plotted using logarithmic scale. The data from this analysis tells us the following: (1) The average scatter is mostly flat at the level of $\sim10^{-7}$ $\mathrm{str}^{-1}$. (2) The low BRDF measured here indicates that there are almost equal scatterers in the bright and dark images, suggesting we are mostly at the noise floor of our measurement. (3) We observe that there is a slow growth of scatter towards the end of the run which may indicate structural changes in the material. However, this requires further investigations and characterization using x-ray crystallography and Raman measurements. 

\section{Conclusion}
In this paper, we studied the temperature induced scatter in a full stack of HR coatings comprised of silica (low n) and titania doped with germania (high n). The results presented here show that in a small region of interest the median BRDF is as low as $1.3 \times 10^{-7}\,\mathrm{str}^{-1}$, a level that would make this coating a viable option for future upgrades to LIGO and Virgo. Newer samples, also from LMA, with slight modifications to the coating recipe have shown to perform better already and investigations are ongoing. Future work in this experiment includes taking into account the beam profile in the analysis to more accurately predict the scatter in a given region of interest, and potentially expanding the study to track scatterers.

\section*{Acknowledgments}
This work was supported by NSF awards 2207998, 2219109 and 2309200 and by Dan Black and Family and Nicholas and Lee Begovich. The authors extend their gratitude to P. Fritschel, M. Fazio, N. Demos, G. Vajente, G. McGhee and the AdV+/A+ Working Group for useful discussions. This paper has LIGO Document No. P2500512.

\end{document}